\documentstyle[12pt,epsf]{article}
\catcode`\@=11
\def \thesection {\arabic{section}.}

\def \be  {\begin{equation}}
\def \ee  {\end{equation}}
\def \ba  {\begin{eqnarray}}
\def \ea  {\end{eqnarray}}
\def \baa {\begin{eqnarray*}}
\def \eaa {\end{eqnarray*}}
\def \bb  {\begin{thebibliography}}
\def \eb  {\end{thebibliography}}
\def \lab #1 {\label{#1}}
\newcommand \ci [1] {\cite{#1}}
\newcommand \bi [1] {\bibitem{#1}}
\newcommand\re[1]{(\ref{#1})}
\def \qqquad {\qquad\quad}

\def \matrix #1 {\left(\begin{array}{cc} #1 \end{array}\right)}

\def \Re {\mathop{\rm Re}\nolimits}

\def \e  {\mathop{\rm e}\nolimits}

\newcommand\lr[1]{{\left({#1}\right)}}

\newcommand \vev [1] {\langle{#1}\rangle}

\newcommand{\as}{\ifmmode\alpha_{\rm s}\else{$\alpha_{\rm s}$}\fi}

\def \CO {{\cal O}}

\font\cmss=cmss12 
\def\inbar{\,\vrule height1.5ex width.4pt depth0pt}
\def\IC{\relax\hbox{$\inbar\kern-.3em{\rm C}$}}
\def\IZ{\relax{\hbox{\cmss Z\kern-.4em Z}}}
\def\IR{{\hbox{{\rm I}\kern-.2em\hbox{\rm R}}}}

\def\IP{{\hbox{{\rm I}\kern-.2em\hbox{\rm P}}}}
\def\II{\hbox{{1}\kern-.25em\hbox{l}}}

\def\numberbysection{\@addtoreset{equation}{section}
                     \def\theequation{\thesection\arabic{equation}}}
\numberbysection

\begin{document}

\def\thefootnote{\fnsymbol{footnote}}
\thispagestyle{empty}

\hfill\parbox{50mm}{{\sc LPTHE--Orsay--97--62} \par
                         hep-ph/9711277        \par
                         November, 1997}
\vspace*{35mm}
\begin{center}
{\LARGE Conformal bootstrap for the BFKL Pomeron}
\par\vspace*{15mm}\par
{\large G.~P.~Korchemsky}
\par\bigskip\par\medskip
{\em Laboratoire de Physique Th\'eorique et Hautes Energies%
\footnote{Laboratoire associ\'e au Centre National de la Recherche
Scientifique (URA D063)} \\
Universit\'e de Paris XI, Centre d'Orsay, b\^at. 211\\
91405 Orsay C\'edex, France}
\par\medskip
{\em and}
\par\medskip
{\em Laboratory of Theoretical Physics,\\
Joint Institute for Nuclear Research, \\
141980 Dubna, Russia}
\end{center}

\vspace*{20mm}

\begin{abstract}
We calculate the interaction vertex of three BFKL states including 
the bare triple BFKL Pomeron coupling and discuss its relation 
with the correlation functions in two-dimensional conformal field 
theory. We construct the operator algebra of the fields interpolating 
the BFKL states and show that in the multi-color limit the vertex 
satisfies the constraints imposed by the conformal bootstrap on the
structure constants of the operator product expansion in conformal 
field theory.
\end{abstract}

\newpage

\def\thefootnote{\arabic{footnote}}
\setcounter{footnote} 0

\section{Introduction}

Recent studies of the high-energy scattering of heavy onium states
within the BFKL approach indicate an intrinsic relation between
the Regge asymptotics of scattering amplitudes in perturbative QCD and
two-dimensional conformal field theories \ci{L}.

The BFKL pomeron appears in the leading logarithmic approximation
as a color-singlet compound state of two interacting reggeized gluons,
or reggeons, propagating in the $t-$channel. The BFKL equation for the
compound reggeon states takes the form of the Schr\"odinger equation
with the hamiltonian ${\cal H}_{\rm BFKL}$ acting on the 2-dimensional
transverse coordinates of the reggeons $\rho=(\rho_x,\rho_y)$. One of the
remarkable properties of the BFKL kernel ${\cal H}_{\rm BFKL}$ is that
it is invariant under projective $SL(2,\IC)$ transformations of the
holomorphic and antiholomorphic reggeon coordinates
\be
z=\rho_x+i\rho_y\,,\qquad
\bar z=z^* \,,\qquad
z\to \frac{az+b}{cz+d} \,,\qquad ad-bc=1\,.
\lab{SL2}
\ee
This property allows to classify the solutions of the BFKL equation
according to the (unitary) principal series representations of the
$SL(2,\IC)$ group and express the wave function
$E_{h,\bar h}(\rho_{10},\rho_{20})$ of the compound state of two
reggeons with the coordinates $\rho_1$ and $\rho_2$ in the form
of three-point correlation function in 2-dimensional conformal
field theory \ci{L}
\be
E_{h,\bar h}(\rho_{10},\rho_{20})=
\lr{\frac{z_{12}}{z_{10}\rho_{20}}}^h
\lr{\frac{\bar z_{12}}{\bar z_{10}\bar z_{20}}}^{\bar h}
=\vev{\phi_{0,0}(z_1,\bar z_1)\phi_{0,0}(z_2,\bar z_2)O_{h,\bar h}(z_0,\bar z_0)}
\,,
\lab{E}
\ee
with $\rho_{jk}=\rho_j-\rho_k$ and
$\rho_0$ being the center-of-mass coordinate of the compound state.
The operators $\phi_{0,0}(z_1,\bar z_1)$ and $\phi_{0,0}(z_2,\bar z_2)$
represent reggeized gluons and have zero conformal weights. One associates
the 2-reggeon compound state with the local operator
$O^{(2)}=O_{h,\bar h}(z_0,\bar z_0)$ carrying the quantum numbers of the
state. The conformal weights $h$ and $\bar h$ of the BFKL state take the
form
\be
h=\frac{1+n}2+i\nu\,,\qquad \bar h=1-h^*
\,,\qquad n=\IZ\,,\quad \nu=\IR\,,
\lab{h}
\ee
where $h-\bar h=n$ and $h+\bar h=1+2i\nu$ are integer conformal
spin and scaling dimension of the state, respectively. It follows from
\re{E} that under $SL(2,\IC)$ transformations of the coordinates \re{SL2}
the operator $O_{h,\bar h}(z,\bar z)$ is transformed as
\be
O_{h,\bar h}(z_0,\bar z_0) \to (cz_0+d)^{2h} (\bar c\bar z_0+\bar d)^{2\bar h}
O_{h,\bar h}(z_0,\bar z_0)
\lab{O-tr}
\ee
and it can be identified \ci{Qua} as a quasiprimary field in two-dimensional 
conformal field theory \ci{BPZ}. The BFKL pomeron appears as the state 
\re{E} with the quantum numbers $h=\bar h=\frac12$ corresponding to the 
ground state energy of the hamiltonian $-{\cal H}_{\rm BFKL}$.

The Regge asymptotics of the scattering amplitude of two onia with
mass $M$ and large center-of-mass energy $s$ is controlled by the
propagator of 2-reggeon compound states, which has the following form
in the impact parameter space \ci{L}
\be
f_Y(\rho_1,\rho_2;\rho_1',\rho_2')=
\sum_{h\neq 0,1} \e^{Y \omega(h)}
\left|\frac{h-\frac12}{h(h-1)}\right|^2
G_{h,\bar h}(\rho_1,\rho_2;\rho_1',\rho_2')
\lab{Phi}
\ee
with $Y=\ln \frac{s}{M^2}\gg 1$.
Here, the summation  goes over all BFKL states \re{E}
except of two degenerate states with the conformal weights $(h=1,\bar h=0)$
and $(h=0,\bar h=1)$. The sum over $h$ is understood as an integral over
continuous $-\infty<\nu<\infty$ and sum over integer spins $n$. The
energy of the BFKL state is defined as an eigenvalue of the BFKL Hamiltonian
${\cal H}_{\rm BFKL}$
\be
\omega(h)=\frac{2\as N_c}{\pi}\Re\left[\psi(1)-\psi(h)\right]\,,
\lab{omega}
\ee
with $\psi(h)=\frac{d}{dx}\ln \Gamma(h)$. The Green function
$G_{h,\bar h}(\rho_1,\rho_2;\rho_1',\rho_2')$ depends on the anharmonic
ratios of the reggeon coordinates $x=\frac{z_{12}z_{1'2'}}{z_{11'}z_{22'}}$
and $\bar x=x^*$ and it can be expressed in terms of the hypergeometric
functions as
\ba
G_{h,\bar h}(\rho_1,\rho_2;\rho_1',\rho_2')&=&
\int d^2\rho_0 E^*_{h,\bar h}(\rho_{1'0},\rho_{2'0})
E_{h,\bar h}(\rho_{10},\rho_{20})
\nonumber
\\
&=&\frac{x^h \bar x^{\bar h}}{B(1-h)} F(h,h;2h;x)
F(\bar h,\bar h;2\bar h;\bar x) + (h\rightleftharpoons 1-h)
\lab{G}
\ea
where $\bar h=1-h^*$ and the function $B(h)$ is defined below in \re{b}.

The scattering amplitude $f_Y(\rho_1,\rho_2;\rho_1',\rho_2')$ is 
invariant under projective
transformations of the reggeon coordinates \re{SL2} and apart from the
$Y-$dependent factor it resembles an expansion of the 4-point correlation
function of reggeon fields $\phi_{0,0}$ over the conformal blocks \ci{BPZ}
in the channel $1 + 2\to 1'+ 2'$. However, this identification is not
valid since $f_Y(\rho_1,\rho_2;\rho_1',\rho_2')$ does not satisfy the 
condition of the crossing symmetry imposed on the 4-point correlation 
functions in conformal field theory.

To preserve unitarity of the scattering amplitudes at very large
energies one has to supplement the BFKL pomeron with unitarity
corrections associated with the diagrams containing an arbitrary
number of reggeized gluons in the $t-$channel \ci{BKP}. These corrections
can be implemented in two steps.

At the first step, one restores the unitarity of the scattering amplitudes
only in the direct channels by taking into account the diagrams with a
conserved number $n\ge 2$ of reggeized gluons in the $t-$channel.
In this case, similar to the BFKL pomeron, the high-energy asymptotics
is controlled by the $n-$reggeon compound states, which satisfy the
Schr\"odinger equation with the Hamiltonian describing a pair-wise
interaction of $n$ reggeons through the BFKL kernel \ci{BKP}. The
$n-$reggeon Schr\"odinger equation inherits the $SL(2,\IC)$ symmetry of the
BFKL equation and, moreover, in the multicolor limit, $N_c\to\infty$, it
acquires additional hidden symmetry due to the fact that the $n$ reggeon
Hamiltonian becomes equivalent to the completely integrable XXX Heisenberg
magnet model \ci{FK,LL}. This symmetry becomes large enough for the
Schr\"odinger equation to be solved exactly by means of the Quantum Inverse
Scattering Method \ci{FK,Qua}. In particular, similar to the BFKL states,
one can construct the quasiprimary fields $O^{(n)}_{h,\bar h}(z_0,\bar z_0)$
interpolating the $n-$reggeon states  and define their wave function
as the $(n+1)-$point correlation function of this field with $n$ additional
reggeon fields $\phi(z_k,\bar z_k) \ci{Qua}$.

At the second step, one restores unitarity of the scattering amplitudes
in the subchannels by including the interaction vertices which
change the number of reggeons. The simplest example of such vertex is
the transition kernel from 2 to 4 reggeized gluons $V_{(2,4)}$
calculated in \ci{B,BW}. Remarkably enough, $V_{(2,4)}$ vertex is also
invariant under projective transformations \ci{BLW} and it is natural
to expect that the $SL(2,\IC)$ invariance is the general feature
of all reggeon transition vertices \ci{BE}. Inclusion of reggeon transition
vertices turns quantum mechanical description of the $n-$reggeon compound
states into interacting quantum field theory.

The reggeon number changing transition kernels induce the interaction
between $n-$reggeon compound states and the corresponding interaction
vertices can be calculated by projecting the transition kernels on the
wave functions of the compound states \ci{Lot}. The resulting expressions
depend on the center-of-mass coordinates and the conformal weights of
the states. The $SL(2,\IC)$ invariance of the transition kernels
implies that the properties of the interaction vertices under
projective transformations \re{SL2} are determined by the properties of
the interacting states. This suggests to identify the interaction vertices
with the correlation functions of quasiprimary fields associated with
reggeon compound states. For example, using the kernel
$V_{(2,4)}=V_{(2,4)}(\rho_1',\rho_2';\rho_1,\rho_2,\rho_3,\rho_4)$
one can calculate the coupling of three BFKL states with
the center-of-mass coordinates $\rho_\alpha$, $\rho_\beta$,
$\rho_\gamma$ and the conformal weights $(h_\alpha,\bar h_\alpha)$,
$(h_\beta,\bar h_\beta)$, $(h_\gamma,\bar h_\gamma)$,
as
\ba
V(\alpha\to\beta,\gamma)
&=&
\int \prod_{{j=1',2'\atop
1,2,3,4}}d^2\rho_j
\
V_{(2,4)}(\rho_1',\rho_2';\rho_1,\rho_2,\rho_3,\rho_4)
\nonumber
\\
&\times&
E_{h_\alpha\bar h_\alpha}(\rho_{1'\alpha},\rho_{2'\alpha})
E_{h_\beta\bar h_\beta}(\rho_{1\beta},\rho_{2\beta})
E_{h_\gamma\bar h_\gamma}(\rho_{3\gamma},\rho_{4\gamma})\,,
\lab{V3}
\ea
where integration goes over the coordinates of reggeons entering the BFKL states
and the result should be compared with the following 3-point correlation
function
\be
V_{\rm CFT}(\alpha,\beta,\gamma)=\vev{
O_{h_\alpha,\bar h_\alpha}(z_\alpha,\bar z_\alpha)
O_{h_\beta,\bar h_\beta}(z_\beta,\bar z_\beta)
O_{h_\gamma,\bar h_\gamma}(z_\gamma,\bar z_\gamma)
}\,.
\lab{CFT}
\ee
The same transition kernel $V_{(2,4)}$ can be also projected on
the wave functions of $n=2$ (or BFKL state) and $n=4$-reggeon compound
states defining the nondiagonal 2-point correlation function
$$
\vev{O^{(2)}_{h_\alpha,\bar h_\alpha}
(z_\alpha,\bar z_\alpha)O^{(4)}_{h_\beta,\bar h_\beta}
(z_\beta,\bar z_\beta)}\,.
$$

It is well known that in conformal field theory the functions \re{CFT}
determine the structure constants of the operator algebra of quasiprimary
fields and they satisfy the set of constraints followed from the
associativity condition of the operator algebra \ci{P,BPZ}.
Therefore, identification of the transition vertices \re{V3} as 3-point
correlation functions \re{CFT} becomes nontrivial since it
imposes severe restrictions on the possible form of the vertex
$V(\alpha\to\beta,\gamma)$. In the present paper we study
the relation between two different representations \re{V3} and \re{CFT} of
the transition vertex of three BFKL states. We calculate the explicit form
of the vertex $V(\alpha\to\beta,\gamma)$ and show that in the multi-color
limit, $N_c\to\infty$ and $\as N_c={\rm fixed}$, it satisfies the
constraints imposed by the conformal bootstrap \ci{P,BPZ}.

The paper is organized as follows. In Sect.~2 we define the vertex
$V(\alpha\to\beta,\gamma)$ and consider its general properties.
The results of calculations are summarized at the end of the section.
In Sect.~3 we construct the operator product expansion of the operators
$O_{h,\bar h}(z,\bar z)$ and identify $V(\alpha\to\beta,\gamma)$
in the multi-color limit as the structure constants of
the operator algebra. Appendices A and B contain the details of
the Feynman diagram techniques used for calculation of the vertex
in Sect.~2.

\section{Transition vertex}

Let us consider the vertex \re{V3} and replace the transition kernel
$V_{(2,4)}$ by its explicit expression in the configuration space
\ci{BLW,Lot}. Straightforward calculation gives the following result \ci{Lot}
\ba
V(\alpha\to\beta,\gamma)&=&\lr{\frac{\as N_c}{\pi}}^2 16
h_\alpha(1-h_\alpha) \bar h_\alpha(1-\bar h_\alpha)
\nonumber
\\
&\times&
\left[V_0(\alpha,\beta,\gamma) - \frac{2\pi}{N_c^2}
V_1(\alpha,\beta,\gamma)
\Re\left\{\psi(1)+\psi(h_\alpha)-\psi(h_\beta)-\psi(h_\gamma)\right\}
\right]\,,
\lab{V}
\ea
where the conformal weights of three BFKL states, $h_\alpha$, $h_\beta$
and $h_\gamma$, are of the form \re{h}.
Each pair of the reggeons coming out of three BFKL wave functions in \re{V3}
is in the color singlet state and according to the color flow inside the
transition kernel one gets planar, $V_0$, and nonplanar, $V_1$, contributions.

In the multi-color limit, only the first term survives in \re{V}
and it has the following form
\be
V_0(\alpha,\beta,\gamma)
=\int\frac{d^2\rho_0d^2\rho_1d^2\rho_2}{|\rho_{01}\rho_{12}\rho_{20}|^2}
E_{h_\alpha\bar h_\alpha}
(\rho_{0\alpha},\rho_{1\alpha})
E_{h_\beta\bar h_\beta}
(\rho_{1\beta},\rho_{2\beta})
E_{h_\gamma\bar h_\gamma}
(\rho_{2\gamma},\rho_{0\gamma})\,,
\lab{A}
\ee
where in the l.h.s.\ the label $\alpha$ denotes the set of conformal
weights $(h_\alpha,\bar h_\alpha)$ and center-of-mass coordinate
$\rho_\alpha$ of the BFKL state. Replacing the BFKL wave functions by
their explicit expressions \re{E} one can represent $V_0$ as a planar
2-dimensional Feynman diagram shown in Fig.~1a. It is interesting to
notice that the same integral contributes to the triple-dipole vertex
\ci{RP} in the QCD dipole model \ci{M}. The nonplanar term in \re{V} is
suppressed by the color factor $1/N_c^2$. It is given by
\be
V_1(\alpha,\beta,\gamma)
=\int\frac{d^2\rho_0d^2\rho_1}{|\rho_{01}|^4}
E_{h_\alpha\bar h_\alpha}
(\rho_{0\alpha},\rho_{1\alpha})
E_{h_\beta\bar h_\beta}
(\rho_{0\beta},\rho_{1\beta})
E_{h_\gamma\bar h_\gamma}
(\rho_{0\gamma},\rho_{1\gamma})
\lab{B}
\ee
and it can be represented as a nonplanar 2-dimensional
Feynman diagram shown in Fig.~1b.

\begin{figure}[ht]
\centerline{\epsffile{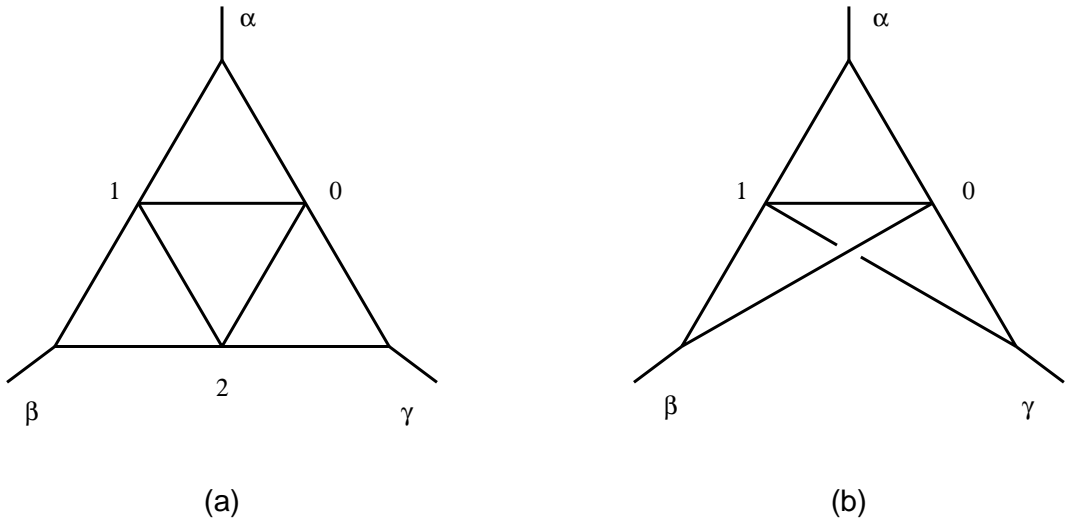}}
\caption{The Feynman diagrams corresponding to the planar (a) and nonplanar
(b) contribution to the vertex $V(\alpha\to\beta,\gamma)$. Solid lines
represent two-dimensional propagators $\frac1{z^h\bar z^{\bar h}}$ with
the exponents $h$ and $\bar h$ depending on the line and the vertices
do not bring any additional factors.}
\end{figure}

One checks that the functions \re{A} and \re{B} are transformed
under $SL(2,\IC)$ transformations in the same way as the
3-point correlation function of quasiprimary fields, \re{CFT} and
\re{O-tr}. This property allows to restore the coordinate dependence of
$V_0$ and $V_1$
\ba
V_0(\alpha_1,\alpha_2,\alpha_3) &=& \Omega(h_1,h_2,h_3) \times
\prod_{i<j} (z_i-z_j)^{-\Delta_{ij}}
(\bar z_i-\bar z_j)^{-\bar\Delta_{ij}}
\nonumber
\\
V_1(\alpha_1,\alpha_2,\alpha_3) &=& \Lambda(h_1,h_2,h_3) \times
\prod_{i<j} (z_i-z_j)^{-\Delta_{ij}}
(\bar z_i-\bar z_j)^{-\bar\Delta_{ij}}
\lab{Omega}
\ea
where $\Omega$ and $\Lambda$ depend only on the
conformal weights $h_j$ and $\bar h_j$ $(j=1,2,3)$ and
\be
\Delta_{12}=h_1+h_2-h_3\,, \qquad
\bar\Delta_{12}=\bar h_1+\bar h_2-\bar h_3\,,
\lab{Deltas}
\ee
etc. Here and in what follows, we indicate explicitly the dependence
of $\Omega$ and $\Lambda$ only on the holomorphic conformal weights
$h_\alpha$, $h_\beta$ and $h_\gamma$, keeping in mind the relation
between holomorphic and antiholomorphic conformal weights, $\bar h=1-h^*$.

Substituting \re{Omega} into \re{V} one finds that the vertex
$V(\alpha\to\beta,\gamma)$ has similar dependence on the center-of-mass
coordinates and its dependence on the conformal weights is given by
\ba
V(h_\alpha;h_\beta,h_\gamma)&=&16\lr{\frac{\as N_c}{\pi}}^2
|h_\alpha(h_\alpha-1)|^2
\lab{Vm}
\\
&\times&
\left(\Omega(h_\alpha,h_\beta,h_\gamma) - \frac{2\pi}{N_c^2}
\Lambda(h_\alpha,h_\beta,h_\gamma)
\Re\left[\psi(1)+\psi(h_\alpha)-\psi(h_\beta)-\psi(h_\gamma)\right]
\right)\,.
\nonumber
\ea
In this expression, $h_\alpha$ is the conformal weight of the incoming
BFKL state and the outgoing states have the conformal weights $h_\beta$
and $h_\gamma$.

\subsection{Symmetry properties of the vertex}

Let us consider some general properties of the vertex
$V(\alpha\to\beta\gamma)$ which become useful for its calculation.
It follows from the definitions \re{A} and \re{B} as well as from the
symmetry of the diagrams in Fig.~1 that the function $V_0$
is invariant under cyclic permutations of its arguments,
while $V_1$ is a completely symmetric function. Then,
using the symmetry property of the BFKL wave function \re{E}
\be
E_{h_\alpha\bar h_\alpha}(\rho_{0\alpha},\rho_{1\alpha})
=(-)^{n_\alpha} E_{h_\alpha\bar h_\alpha}(\rho_{1\alpha},\rho_{0\alpha})
\lab{sym}
\ee
it is easy to show from \re{Omega} that $\Omega$ and $\Lambda$ are
symmetric functions of the conformal weights of the BFKL states
\be
 \Omega(h_\alpha,h_\beta,h_\gamma)
=\Omega(h_\beta,h_\alpha,h_\gamma)
=\Omega(h_\gamma,h_\beta,h_\alpha)\,.
\lab{s}
\ee
The function $\Lambda$ satisfies similar relations plus additional
selection rule on the conformal spins
of three BFKL states, $n_\alpha$, $n_\beta$ and $n_\gamma$,
\be
\Lambda(h_\alpha,h_\beta,h_\gamma)=0 \qquad
{\rm for} \quad n_\alpha+n_\beta+n_\gamma={\rm odd}\,,
\lab{rule}
\ee
which one obtains by changing the integration
variables in \re{B} as $\rho_0\rightleftharpoons\rho_1$ and applying
the identity \re{sym}.

Another set of identities follows from the intertwining relation
for the BFKL wave function \ci{L}
\be
\lr{E_{h,\bar h}(\rho_{10},\rho_{20})}^*
=E_{1-h,1-\bar h}(\rho_{10},\rho_{20})
=B(h) \int d^2\rho_{0'}\, E_{h,\bar h}(\rho_{10'},\rho_{20'})\,
z_{00'}^{2(h-1)} \bar z_{00'}^{2(\bar h-1)}\,,
\lab{inter}
\ee
which expresses the fact that the principal series representations
of the $SL(2,\IC)$ group labelled by the conformal weights $(h,\bar h)$
and $(1-h,1-\bar h)$ are unitary equivalent. Here, the prefactor is
given by  
\be
B(h)=\frac{(-)^n}{\pi}\frac{\Gamma^2(\bar h)}{\Gamma^2(1-h)}
\frac{\Gamma(2-2h)}{\Gamma(2\bar h -1)}
=
\frac{(-)^n}{\pi}2^{-4i\nu} (n-2i\nu)
\frac{\Gamma(\frac{1+n}2+i\nu)\Gamma(\frac{n}2-i\nu)}
{\Gamma(\frac{1+n}2-i\nu)\Gamma(\frac{n}2+i\nu)}\,.
\lab{b}
\ee
As a result, one gets from \re{A} and \re{inter}
\be
\lr{{}^{}\Omega(h_\alpha,h_\beta,h_\gamma)}^*
=\Omega(1-h_\alpha,1-h_\beta,1-h_\gamma)
\lab{cc}
\ee
and similar relation for the function $\Lambda$.
Let us multiply the both sides of \re{A} by
$z_{\gamma\gamma'}^{2(h_\gamma-1)}
\bar z_{\gamma\gamma'}^{2(\bar h_\gamma-1)}$ and integrate
with respect to the center-of-mass coordinate $\rho_\gamma$.
Then, one applies the identity \re{inter} in the r.h.s., performs
2-dimensional integration in the l.h.s.\  and replaces the function
$V_0$ by its expression \re{Omega} to arrive at the following relation
\be
\frac{\Omega(h_\alpha,h_\beta,1-h_\gamma)}{\Omega(h_\alpha,h_\beta,h_\gamma)}
=(-)^{n_\alpha+n_\beta}
\frac{\Gamma^2(1-h_\gamma^*)}{\Gamma^2(1-h_\gamma)}
\frac{\Gamma(1+h_\alpha-h_\beta-h_\gamma)\Gamma(1-h_\alpha+h_\beta-h_\gamma)}
{\Gamma(1+h_\alpha^*-h_\beta^*-h_\gamma^*)
           \Gamma(1-h_\alpha^*+h_\beta^*-h_\gamma^*)}\,.
\lab{O}
\ee
Here, the ratio of $\Gamma-$functions is a pure phase and it vanishes
for real values of the conformal weights. Let us consider two special
cases of \re{O}
$$
\frac{\Omega(h_\alpha,h_\alpha,1-h_\gamma)}
     {\Omega(h_\alpha,h_\alpha,h_\gamma)}=1\,,
\qqquad
\frac{\Omega(h_\alpha,h_\beta,1)}
     {\Omega(h_\alpha,h_\beta,0)}=
     \frac{h_\alpha-h_\beta}{h_\alpha^*-h_\beta^*}\,.
$$
The same transformations been applied to $V_1$ lead to the
same set of identities for the function $\Lambda$.

Let us show that the functions $V_0$ and $V_1$ satisfy certain bilinear
relations. The latter follow from the completeness condition for the BFKL
states \ci{L}
\be
\sum_{h} d(h) \int d^2\rho_0\, E_{h,\bar h}(\rho_{10},\rho_{20})
E_{1-h,1-\bar h}(\rho_{1'0},\rho_{2'0})
=|\rho_{12}|^2 |\rho_{1'2'}|^2 \,\delta^2(\rho_{11'})\delta^2(\rho_{22'})\,,
\lab{comp}
\ee
where integration goes over the center-of-mass coordinate and the summation
over conformal weights, \re{h}, goes as in \re{Phi}. The prefactor $d(h)$
is related to the norm of the BFKL state and it is given by
\be
d(h)
=\frac1{\pi^4} \left|h-\mbox{$\frac12$}\right|^2
=\frac1{(2\pi)^2} |B(h)|^2
=\frac1{(2\pi)^2} B(h) B(1-\bar h)\,,
\lab{norm}
\ee
with the function $B(h)$ defined in \re{b}. To apply \re{comp}
one identifies two BFKL wave functions entering the l.h.s.\ of \re{comp}
as belonging to two different functions $V_0$. This leads to the
following expression for the product of two functions $V_0$
summed over the quantum numbers of the ``intermediate'' BFKL states%
\footnote{The integral entering the r.h.s.\ of this relation
can be interpreted as the quadruple dipole vertex in the QCD dipole model 
\ci{RP}.}
\ba
&&\hspace*{-10mm}
\sum_{h_\gamma} d(h_\gamma) \int d^2 \rho_\gamma \, V_0(\alpha,\beta,\gamma)
V_0(1-\gamma,\beta',\alpha')
\lab{4}
\\
&=&\int \frac{d^2\rho_0d^2\rho_1d^2\rho_2d^2\rho_3}
{|\rho_{01}\rho_{12}\rho_{23}\rho_{30}|^2}
E_{h_\alpha\bar h_\alpha}
(\rho_{0\alpha},\rho_{1\alpha})
E_{h_\beta\bar h_\beta}
(\rho_{1\beta},\rho_{2\beta})
E_{h_\beta'\bar h_\beta'}
(\rho_{2\beta'},\rho_{3\beta'})
E_{h_\alpha'\bar h_\alpha'}
(\rho_{3\alpha'},\rho_{0\alpha'})\,.
\nonumber
\ea
Here, the label $1-\gamma$ denotes the BFKL state with the conformal weights
$(1-h_\gamma,1-\bar h_\gamma)$ and the center-of-mass coordinate
$\rho_\gamma$.
We notice that the r.h.s.\ of this relation
is invariant under cyclic permutations of the BFKL states $\alpha$,
$\beta$, $\beta'$ and $\alpha'$ leading to the following
crossing symmetry property
\be
\sum_{h_\gamma} d(h_\gamma) \int d^2\rho_\gamma\, V_0(\alpha,\beta,\gamma)
V_0(1-\gamma,\beta',\alpha')
=
\sum_{h_\gamma} d(h_\gamma) \int d^2\rho_\gamma\, V_0(\alpha',\alpha,\gamma)
V_0(1-\gamma,\beta,\beta')\,.
\lab{cr}
\ee
The relations \re{4} and \re{cr} can be depicted as shown in Fig.~2.
\begin{figure}[ht]
\centerline{\centerline{\epsffile{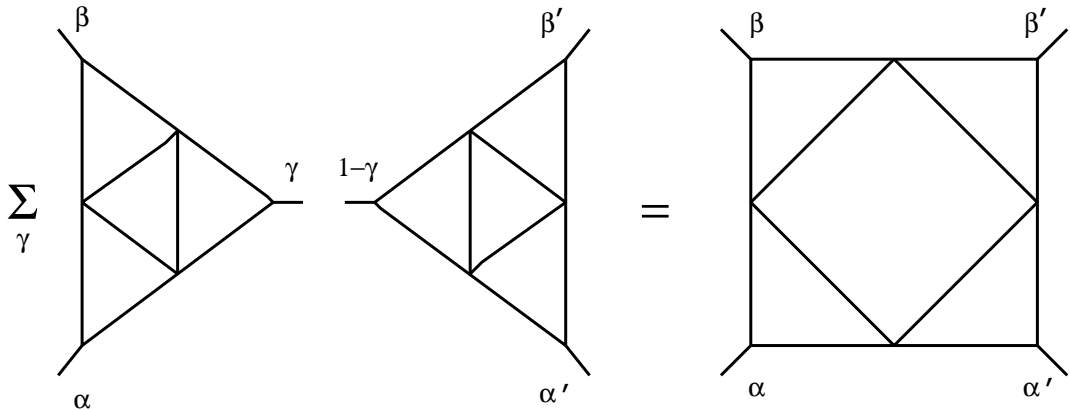}}}
\caption{Graphical form of the relation \re{4}. The crossing
property \re{cr} follows from the symmetry of the diagram in
the r.h.s.
}
\end{figure}
Repeating similar analysis one can show that the function $V_1$ also
satisfies the same relation and at the same time it does not hold for the
cross product of the functions $V_0$ and $V_1$. As we will show in
Sect.~3.2 the bilinear relations \re{cr} give rise to the associativity
condition of the operator algebra of the quasiprimary fields $O_{h,\bar h}$.

\subsection{Nonplanar diagram}

Let us start with the nonplanar contribution to the interaction
vertex given by \re{B}. Its calculation is simpler
than that of the planar diagram, \re{A}. Using the $SL(2,\IC)$
symmetry one chooses the center-of-mass coordinates of three BFKL
states at
\be
z_\alpha=\bar z_\alpha=0\,,\qquad
z_\beta=\bar z_\beta=1\,,\qquad
z_\gamma=\bar z_\gamma=\infty
\lab{mass}
\ee
and obtains from \re{B} and \re{Omega}
the function $\Lambda$ as the following 2-loop integral
\be
\Lambda(h_\alpha,h_\beta,h_\gamma)=
\int\frac{d^2z_0d^2z_1}{|z_{01}|^4}
\lr{\frac{z_{01}}{z_0z_1}}^{h_\alpha}
\lr{\frac{\bar z_{01}}{\bar z_0\bar z_1}}^{\bar h_\alpha}
\lr{\frac{z_{01}}{(1-z_0)(1-z_1)}}^{h_\beta}
\lr{\frac{\bar z_{01}}{(1-\bar z_0)(1-\bar z_1)}}^{\bar h_\beta}
z_{01}^{h_\gamma}\bar z_{01}^{\bar h_\gamma}
\lab{2-loop}
\ee
where $h_\alpha=\frac{1+n_\alpha}2+i\nu_\alpha$, $\bar h_\alpha=1-h_\alpha^*$
etc. In this expression, $z=\rho_x+i\rho_y$ and $\bar z=\rho_x-i\rho_y$
are complex valued holomorphic and antiholomorphic reggeon coordinates
and $d^2z=dxdy=\frac12dzd\bar z$.

We present here the final result of the calculation of the integral \re{2-loop}
and refer to the Appendix A for the details. Applying the techniques described
in Appendix A, one can express $\Lambda$ as a sum of two terms each given
by the product of two contour integrals, $J$ and $\bar J$, over holomorphic
and antiholomorphic coordinates, respectively. These two terms differ only
by a sign factor $(-)^{n_\alpha+n_\beta+n_\gamma}$ and their sum can be
represented as
\be
\Lambda(h_\alpha,h_\beta,h_\gamma)
=
\frac{\pi^2 \left[{1+(-)^{n_\alpha+n_\beta+n_\gamma}}\right]
}
{\Gamma^2(h_\alpha)\Gamma^2(h_\beta)\Gamma(2-h_\alpha-h_\beta-h_\gamma)}
\
J(h_\alpha,h_\beta,h_\gamma)
\times
\bar J(\bar h_\alpha,\bar h_\beta,\bar h_\gamma)\,.
\lab{form}
\ee
In this form, $\Lambda$ explicitly satisfies the selection rule \re{rule}.
As shown in Appendix A, the integrals $J$ and $\bar J$ are given by
the ratio of $\Gamma-$functions leading to the following result
\ba
\Lambda(h_\alpha,h_\beta,h_\gamma)
&=&{\pi^2}\cos\lr{\frac{\pi}2(n_\alpha+n_\beta+n_\beta)}
2^{h_\alpha+\bar h_\alpha+h_\beta+\bar h_\beta+h_\gamma+\bar h_\gamma-4}
\nonumber
\\
&\times&
\frac{\Gamma(1-\bar h_\alpha)}{\Gamma(h_\alpha)}
\frac{\Gamma(1-\bar h_\beta)}{\Gamma(h_\beta)}
\frac{\Gamma(1-\bar h_\gamma)}{\Gamma(h_\gamma)}
\frac{\Gamma(-\frac12+\frac12(\bar h_\alpha+\bar h_\beta+\bar h_\gamma))}
{\Gamma(\frac32-\frac12(h_\alpha+h_\beta+h_\gamma))}
\lab{Lam}
\\
&\times&
\frac{\Gamma(\frac12(\bar h_\alpha+\bar h_\beta-\bar h_\gamma))}
{\Gamma(1-\frac12(h_\alpha+ h_\beta- h_\gamma))}
\frac{\Gamma(\frac12(-\bar h_\alpha+\bar h_\beta+\bar h_\gamma))}
{\Gamma(1-\frac12(- h_\alpha+ h_\beta+ h_\gamma))}
\frac{\Gamma(\frac12(\bar h_\alpha-\bar h_\beta+\bar h_\gamma))}
{\Gamma(1-\frac12( h_\alpha- h_\beta+ h_\gamma))}\,.
\nonumber
\ea
One verifies that this expression satisfies the relations
\re{s}, \re{rule} and \re{O}. We recall that
$(h_\alpha,\bar h_\alpha)$, $(h_\beta,\bar h_\beta)$ and
$(h_\gamma,\bar h_\gamma)$ are the conformal weights of three
BFKL states defined in \re{h}.

Let us consider the general expression \re{Lam} in two special
cases.
In the first case, one takes $h_\alpha=h$ and $h_\beta=h_\gamma=1$.
We notice that the BFKL states with the conformal weights $h=1$ and
$\bar h=0$ do not contribute to the scattering amplitude \re{Phi}.
As we will argue in Sect.~3.1, according to their conformal properties,
the same states can be formally associated with the states of reggeized 
gluons. Under this identification, the function $\Lambda\lr{h,1,1}$ measures
the nonplanar coupling of two reggeized gluons to the BFKL state
\be
\Lambda\lr{h,1,1}=
\frac{2\pi^2}{\bar h(\bar h-1)} \,,\qquad
\mbox{for $n=$\ even}
\lab{L-zero}
\ee
and it vanishes for odd spins $n$.
In the second case, $h_\alpha=h_\beta=h_\gamma=\frac12$ (and
as a consequence $\bar h_\alpha=\bar h_\beta=\bar h_\gamma=\frac12$)
corresponding to the coupling of three BFKL Pomerons one obtains from \re{Lam}
\be
\Lambda\lr{\mbox{$\frac12,\frac12,\frac12$}}=\frac{2\pi^6}{\Gamma^8(\frac34)}
=378.145
\lab{L1/2}
\ee
This expression determines the nonplanar contribution to the bare triple
BFKL Pomeron vertex \re{Vm}.

\subsection{Planar diagram}

Calculation of the planar diagram follows the same steps as that
of the nonplanar diagram. Choosing the center-of-mass coordinates
in \re{A} and \re{Omega} according to \re{mass} one writes the
function $\Omega$ as the following 3-loop integral
\be
\Omega
=
\int\frac{d^2z_0d^2z_1d^2z_2}{|z_{01}z_{12}z_{20}|^2}
\lr{\frac{z_{01}}{z_0z_1}}^{h_\alpha}
\lr{\frac{\bar z_{01}}{\bar z_0\bar z_1}}^{\bar h_\alpha}
\lr{\frac{z_{12}}{(1-z_1)(1-z_2)}}^{h_\beta}
\lr{\frac{\bar z_{12}}{(1-\bar z_1)(1-\bar z_2)}}^{\bar h_\beta}
z_{20}^{h_\gamma}\bar z_{20}^{\bar h_\gamma}\,.
\lab{Om}
\ee
Its calculation is performed in Appendix B. Separating integrals
over holomorphic and antiholomorphic coordinates one can
express $\Omega$ as a sum of three terms each given by the
product of two contour integrals depending separately on the
holomorphic and antiholomorphic conformal weights
\be
\Omega=\pi^3 \left[\Gamma^2(h_\alpha)\Gamma^2(h_\beta)\Gamma(1-h_\alpha)
\Gamma(1-h_\beta)\Gamma(1-h_\gamma)\right]^{-1}
\ \sum_{a=1}^3 J_a(h_\alpha,h_\beta,h_\gamma)
\times
\bar J_a(\bar h_\alpha,\bar h_\beta, \bar h_\gamma)
\,.
\lab{Ja}
\ee
Explicit expressions for the integrals $J_a$ and $\bar J_a$
can be found in Appendix B. For general values of the conformal weights,
the functions $J_a$ and $\bar J_a$ can be calculated in terms of the
Meijer's $G_{44}^{pq}-$function, which in turn can be expanded over
${}_4F_3-$hypergeometric series and its derivatives with respect to
indices \ci{spec}. Instead of presenting the general expression for the
function $\Omega$, we consider two special physically most interesting
cases.

In the first case, we put $h_\alpha=h_\beta=1$ and $h_\gamma=h$. The
function $\Omega(1,1,h)$ measures the planar coupling of
two reggeized gluons to the BFKL state with the conformal weight
$(h,\bar h)$. Each of the integrals entering \re{Ja} is singular at
$h_\alpha=h_\beta=1$ (or equivalently $\bar h_\alpha=\bar h_\beta=0$)
and one regularizes them by replacing $h_\alpha=1+i\nu_\alpha$ and
$h_\beta=1+i\nu_\beta$ as $\nu_\alpha$, $\nu_\beta\to 0$. In this limit,
the integrals $J_a$ and $\bar J_a$ are given by \re{JaB}.
Substituting them into \re{Ja} one finds that the poles in $\nu_\alpha$
and $\nu_\beta$ are cancelled in the sum of integrals leading to the finite
expression for the interaction vertex%
\footnote{Here we used the symmetry property \re{s},
$\Omega(1,1,h)=\Omega(h,1,1)$.}
\be
\Omega(h,1,1)
=\frac{4\pi^3}{\bar h(1-\bar h)}
\Re\left[\psi(1)-\psi(h)\right]
\,,
\lab{11h}
\ee
which satisfies the relation \re{O}. Taking $h=\frac12$ one calculates the
planar coupling of the BFKL Pomeron to two reggeized gluons as
\be
\Omega\lr{\mbox{$\frac12$},1,1}=32\pi^3\ln 2\,.
\lab{11}
\ee
Let us consider the limit $h=1+i\nu$ as $\nu\to 0$ corresponding to
the coupling of three reggeized gluons. One finds from \re{11h} that
the function $\Omega$ vanishes as
$$
\Omega(1+i\nu,1,1)=4i\pi^3 \zeta(3) \nu + \CO(\nu^2) \,.
$$
Moreover, this case corresponds to the values of conformal spins,
$n_\alpha=n_\beta=n_\gamma=1$, and according to the selection
rules \re{rule}, the nonplanar contribution \re{L-zero}
vanishes identically for any $\nu$. Therefore, substituting \re{L-zero}
into \re{V} one finds that the triple reggeon vertex vanishes as
$$
V (1+i\nu;1,1)\sim \nu^3
$$
with two additional powers of $\nu$ coming from the prefactor in \re{Vm}.
This property of the transition vertex is in agreement with the
Gribov's signature conservation rule which prohibits the existence of
the vertices with odd number of interacting reggeized gluons.
Two reggeons could couple however to the BFKL states with $h\neq 1\,, 0$.
The corresponding transition vertex can be calculated from \re{Vm}, using
\re{L-zero} and \re{11h}, as
$$
V(h;1,1)=\lr{\frac{\as N_c}{\pi}}^2
(4\pi)^3 h(1-h) \lr{1-\frac1{N_c^2}}\Re\left[\psi(1)-\psi(h)\right]\,,
$$
where the nonplanar $1/N_c^2-$term is absent for odd spin $n$.

In the second case, one chooses the conformal weights as
$h_\alpha=h_\beta=h$ and $h_\gamma=\frac12$ and obtains from \re{Ja}
the coupling of the BFKL Pomeron to two BFKL states with the conformal
weights $(h,\bar h)$. The values of integrals $J_a$ and $\bar J_a$
are given by \re{JaP} and their substitution into \re{Ja} yields the
following result
\be
\Omega\lr{\mbox{$\frac12$},h,h}=
-\pi^5\frac{\Gamma(2h-\frac12)\Gamma(2\bar h-\frac12)}
{\Gamma^2(h)\Gamma^2(\bar h)}
\frac{\cot^3(\pi h)}{(h-\frac12)^2(\bar h-\frac12)}\
{}_4F_3(\bar h)\
{}_6F_5(h) + \lr{h\rightleftharpoons \bar h}\,,
\lab{hh1/2}
\ee
where the notations were introduced for the following combinations of
the generalized hypergeometric series
\ba
{}_4F_3(h)&=&{}_4F_3\lr{{{\frac12,\frac12,h,1-h}\atop{1,\frac12+h,\frac32-h}}
\bigg| 1}
\nonumber
\\
{}_6F_5(h)&=&\left[1+4\frac{d}{d x}\right]
{}_6F_5\lr{{{\frac12,\frac12,h,h,1-h,1-h}\atop
{1,\frac12+h,\frac12+h,\frac32-h,\frac32-h}
}\bigg|x=1}
\lab{FF}
\\
&=&{}_6F_5\lr{{{\frac12,\frac12,h,h,1-h,1-h}\atop
{1,\frac12+h,\frac12+h,\frac32-h,\frac32-h}
}\bigg|1}
\nonumber
\\
&+&
{}_6F_5\lr{{{\frac32,\frac32,h+1,h+1,2-h,2-h}\atop
{2,\frac32+h,\frac32+h,\frac52-h,\frac52-h}
}\bigg|1}\frac{(1-h)^{2}h^2}{(\frac12+h)^{2}(\frac32-h)^{2}}\,.
\nonumber
\ea
One verifies that this expression satisfies the relation \re{cc}
and it reproduces \re{11} for $h=1$. Taking $h=\frac12$ one
calculates the planar contribution to the bare triple
BFKL Pomeron vertex as
\be
\Omega\lr{\mbox{$\frac12,\frac12,\frac12$}}
=2\pi^7 {}_4F_3(\mbox{$\frac12$})\ {}_6F_5(\mbox{$\frac12$})
=7766.679
\lab{O1/2}
\ee
This exact value is close to the one \ci{VB} obtained by numerical
Monte Carlo integration of \re{Om}.

Finally, one substitutes \re{L1/2} and \re{O1/2} into \re{Vm} to
obtain the bare triple BFKL Pomeron interaction vertex as
$$
V\lr{\mbox{$\frac12;\frac12,\frac12$}}=
\lr{\frac{\as N_c}{\pi}}^2
\left[
\Omega\lr{\mbox{$\frac12,\frac12,\frac12$}}-
\frac{4\pi\ln 2}{N_c^2} \Lambda\lr{\mbox{$\frac12,\frac12,\frac12$}}
\right]
=\lr{\frac{\as N_c}{\pi}}^2\ 2\pi^7 \lr{1.286- \frac{0.545}{N_c^2}}\,.
$$
We observe that the nonplanar contribution is negative and for $N_c=3$ it
is much smaller than the planar one.

\section{Conformal bootstrap}

Let us interpret the transition vertex calculated in the previous section
as a 3-point correlation function of the quasiprimary fields in two-dimensional
conformal field theory. Following the bootstrap approach \ci{P,BPZ} we will
specify the set of local operators and construct their operator algebra.
The latter will allow us to calculate the correlation function of the
operators \re{CFT} and compare it with the transition vertex \re{V3}.

\subsection{Operator algebra}

Let us associate the quasiprimary operator $O_{h,\bar h}(z,\bar z)$ with
the BFKL state having the conformal weights $(h,\bar h)$
and the center-of-mass coordinate $\rho$. According to the
possible values of the conformal weights, \re{h}, the number of the
operators is infinite. Similarly, one associates the operator
$\phi=\phi_{0,0}(z,\bar z)$ with the reggeized gluons.
Since the BFKL state is built from two reggeized gluons one has
to include the operators $\phi(z,\bar z)$ into
the operator algebra of the fields $O_{h,\bar h}(z,\bar z)$. The operator
$\phi(z,\bar z)$ has to carry the color charge of gluon and in order to
simplify the color structure of the correlation functions involving the
reggeon fields one considers the multi-color limit, $N_c\to \infty$. In
this limit, the color flow becomes simple and one can treat
$\phi(z,\bar z)$ as a scalar operator.

It follows from \re{E} that $\phi_{0,0}(z,\bar z)$
has vanishing conformal weights while in the unitary conformal field
theory the only operator having this property is the identity operator $\II$.
To avoid this problem one notices that taking derivative of the both sides
of \re{E} with respect to $z_1$ and $z_2$ one can express the
BFKL wave function as
\be
E_{h,\bar h}(z_{10},z_{20})=\frac1{h(h-1)}
z_{12}^2
\vev{\partial \phi(z_1,\bar z_1)\partial \phi(z_2,\bar z_2)
O_{h,\bar h}(z_0,\bar z_0)}
\lab{rel}
\ee
where $\partial\phi(z,\bar z)=\frac{\partial}{\partial z}\phi(z,\bar z)$.
This identity allows us to associate the reggeized gluon with the operator
$\partial\phi(z,\bar z)$ having the conformal weights $h=1$ and $\bar h=0$.
These values fit into the spectrum of the conformal weights of the fields
$O_{h,\bar h}(z,\bar z)$ at $n=1$ and $\nu=0$ and one has to distinguish
between the operators $O_{1,0}(z,\bar z)$ and
$\partial\phi_{0,0}(z,\bar z)$ as corresponding to the BFKL state and
reggeized gluon, respectively.

However, examining the BFKL state with $h=1$ and $\bar h=0$ one finds
that its wave function \re{E} is given by the sum of two terms,
$E_{1,0}(\rho_{01},\rho_{02})=\frac1{z_{20}}-\frac1{z_{10}}$, each does
not depending on one of the reggeon coordinates. This implies that the
BFKL state $E_{1,0}(\rho_{01},\rho_{02})$ cannot couple to the gauge
invariant physical states like onium. Indeed, the coupling
is proportional to $\int d^2 \rho_1 d^2 \rho_2 \Phi(\rho_1,\rho_2)
E_{1,0}(\rho_{01},\rho_{02})$ with $\Phi$ being the onium wave
function and it vanishes due to the condition of gauge
invariance \ci{L}, $\int d^2\rho_1\,\Phi(\rho_1,\rho_2)
=\int d^2\rho_2\,\Phi(\rho_1,\rho_2)=0$. This means that the BFKL
state with the conformal weights $(1,0)$ (as well as $(0,1)$) is the
degenerate one and it should be excluded from the spectrum of
physical $n=2$ reggeon compound states, \re{G}. The same property allows us
to identify the corresponding two ``unphysical'' operators,
$O_{1,0}(z,\bar z)$ and $O_{0,1}(z,\bar z)=\lr{O_{1,0}(z,\bar z)}^*$,
as the fields $\partial\phi(z,\bar z)$ and $\bar\partial\phi(z,\bar z)$,
respectively.
Therefore, once we will construct the operator algebra of the
quasiprimary fields $O_{h,\bar h}(z,\bar z)$, the reggeized gluons will be
automatically included into it as the special case $h=1$, $\bar h=0$ and
$h=0$, $\bar h=1$.

The construction of the operator algebra goes as follows.
Let us define the 2-point correlation function of the operators
$O_{h,\bar h}(z,\bar z)$ as
\be
\vev{O_{h_1,\bar h_1}(z_1,\bar z_1) O_{h_2,\bar h_2}(z_2,\bar z_2)}
=D(h_1)\ \delta_{h_1,h_2}\, z_{12}^{-2h_1} \bar z_{12}^{-2\bar h_1}
\lab{2}
\ee
where $\delta_{h_1,h_2}=\delta_{n_1,n_2} \delta(\nu_1-\nu_2)$ and
$D(h)$ is some function to be determined later on. The choice of
$D(h)$ fixes normalization of the operators. Let us define the
3-point correlation function as
\be
\vev{O_{h_1,\bar h_1}(z_1,\bar z_1)
     O_{h_2,\bar h_2}(z_2,\bar z_2)
     O_{h_3,\bar h_3}(z_3,\bar z_3)}
=\Omega(h_1,h_2,h_3) \prod_{i<j} (z_i-z_j)^{-\Delta_{ij}}
(\bar z_i-\bar z_j)^{-\bar \Delta_{ij}}\,,
\lab{3}
\ee
where the function $\Omega$ determines the planar contribution to
the interaction vertex \re{Vm} and the exponents $\Delta_{ij}$ 
and $\bar \Delta_{ij}$ are 
given by \re{Deltas}. Combining together \re{2} and \re{3} we
obtain the following operator algebra for the infinite set of
local operators $\{O_{h,\bar h}(z,\bar z)\,,\II\}$
\ba
O_{h_1,\bar h_1}(z_1,\bar z_1) O_{h_2,\bar h_2}(z_2,\bar z_2)
&=&\II \times D(h_1)\delta_{h_1,h_2}\, z_{12}^{-2h_1} \bar z_{12}^{-2\bar h_1}
\lab{OA}
\\[3mm]
&&\hspace*{-40mm}
+\sum_{h_3} \int d^2 z_3\, O_{1-h_3,1-\bar h_3}(z_3,\bar z_3)
\times
\frac{B(h_3)}{D(1-h_3)}
\Omega(h_1,h_2,h_3) \prod_{i<j} (z_i-z_j)^{-\Delta_{ij}}
(\bar z_i-\bar z_j)^{-\bar \Delta_{ij}}\,.
\nonumber
\ea
Here, the sum goes over the quasiprimary operators with conformal weights
satisfying \re{h}.
Two operators entering the l.h.s.\ of \re{OA} are transformed under
projective transformations according to the principal series
representations of $SL(2,\IC)$ denoted as $t^{(n_1,\nu_1)}$ and
$t^{(n_2,\nu_2)}$. The operator algebra \re{OA} can be interpreted as
the decomposition of the tensor product of two representations over
irreducible components, $t^{(n_1,\nu_1)}\otimes t^{(n_2,\nu_2)}=
\bigoplus_{\nu_3,n_3} t^{(n_3,\nu_3)}$.

The coefficient in front of the operator $O_{1-h_3,1-\bar h_3}
(z_3,\bar z_3)$ in \re{OA} is chosen in such a way that substitution
of \re{OA} into \re{3} reproduces the 3-point function. To perform
this check one applies \re{OA} and integrates over $z_3$ using the
representation \re{Omega} and \re{A} for the function $\Omega$
and by taking into account the intertwining relation \re{inter}.

Two terms entering the r.h.s.\ of \re{OA} have a simple physical
meaning. The first term describes the propagation of the BFKL
state while the second one corresponds to the transition of two BFKL
states into a single BFKL state. The origin of the operator algebra
can be traced back to the projective invariance of the transition vertex
$V_{2,4}$ and assuming that the same property holds for all transition
vertices $V_{n,m}$ one can generalize the relation \re{OA} to take
into account the transitions between the BFKL states and higher $n>2$
reggeon compound states. This can be achieved by including additional
quasiprimary operators $O^{(n)}_{h,\bar h}$ interpolating the $n-$reggeon
compound states into the operator algebra \re{OA}. Their contribution
to the r.h.s.\ of \re{OA} will be however suppressed in the multi-color
limit by the additional factor $(\as N_c)^{n-2}$ and can be considered as
higher order correction to \re{OA}.

\subsection{Crossing symmetry}

The associativity condition of the operator algebra \re{OA} which is
the main dynamical principle of the conformal bootstrap approach \ci{P},
imposes restrictions on the possible form the function $\Omega$ and
it is not obvious that this function defined before as a multi-color
limit of the interaction vertex of three BFKL states, \re{Vm}, satisfies
them. The same condition can be expressed as the crossing symmetry of the
4-point function \ci{BPZ}
\be
V_{\rm CFT}(\alpha_1,\alpha_2,\alpha_3,\alpha_4)=
\vev{O_{h_1,\bar h_1}(z_1,\bar z_1)O_{h_2,\bar h_2}(z_2,\bar z_2)
 O_{h_3,\bar h_3}(z_3,\bar z_3)O_{h_4,\bar h_4}(z_4,\bar z_4)}\,.
\lab{V4}
\ee
Applying \re{OA} together with \re{3} one can write the 4-point function
as a product of two 3-point functions summed over all intermediate states
$\alpha$
$$
V_{\rm CFT}(\alpha_1,\alpha_2,\alpha_3,\alpha_4)=
\sum_{h} \int d^2\rho_\alpha\
V_0(\alpha_1,\alpha_2,\alpha) \frac{B(h)}{D(1-h)}
V_0(1-\alpha,\alpha_3,\alpha_4)\,,
$$
where we used the notations introduced in \re{Omega} and \re{4}.
Comparing this expression with \re{4} one notices that
the 4-point function $V_{\rm CFT}(\alpha,\beta,\beta',\alpha')$
coincides with the r.h.s.\ of \re{4} and,
as a consequence, obeys the crossing symmetry \re{cr} provided that
$$
\frac{B(h)}{D(1-h)}={\rm const}\times d(h)\,.
$$
Substituting \re{norm} into this relation one finds that in order
to satisfy the associativity condition, the 2-point function \re{2}
should be fixed (up to an overall constant factor) as
\be
D(h)=\frac{(2\pi)^2}{B(\bar h)}
\lab{D}
\ee
with the function $B(h)$ defined in \re{b}.

At short distances the operator algebra \re{OA} can be represented in
the form of the local operator product expansion.
Namely, expanding the operator
$O_{1-h_3,1-\bar h_3}(z_3,\bar z_3)$
in powers of $z_{32}$ and $\bar z_{32}$ and performing $z_3-$integration in
\re{OA} one gets
\ba
O_{h_1,\bar h_1}(z,\bar z) O_{h_2,\bar h_2}(0,0)&=&
\II\times D(h_1)\delta_{h_1,h_2}\, z^{-2h_1} \bar z^{-2\bar h_1}
+\sum_{h_3} \frac{\Omega(h_1,h_2,h_3)}{D(h_3)}
z^{-\Delta_{12}} \bar z^{-\bar\Delta_{12}}
\lab{OPE}
\\
&&\times
F(h_1-h_2+h_3;2h_3;z\partial)\,
F(\bar h_1-\bar h_2+\bar h_3;2\bar h_3;\bar z\bar \partial)\
O_{h_3,\bar h_3}(0,0)\,,
\nonumber
\ea
where $F(a;b;x)=1+\frac{a}{b}x +\frac{a+1}{2!(b+1)}x^2 + ...$
denotes the degenerate hypergeometric function. This
relation coincides with the operator expansion of quasiprimary
operators in conformal field theory \ci{BPZ}.

\subsection{Conformal blocks}

The 4-point function of quasiprimary fields, \re{V4}, is given by the
4-fold integral entering the r.h.s.\ of \re{4} and depicted in Fig.~2.
Let us apply the
operator product expansion \re{OPE} to expand it over the conformal blocks.
The function $V_{\rm CFT}(\alpha_1,\alpha_2,\alpha_3,\alpha_4)$
depends on two anharmonic ratios
\be
x=\frac{z_{12}z_{34}}{z_{13}z_{24}}
\,,\qquad
\bar x=x^*
\,.
\lab{x}
\ee
In the standard way one chooses the coordinates as
$z_1=\bar z_1=\infty$, $z_2=\bar z_2=1$, $z_3=x$, $\bar z_3=\bar x$
and $z_4=\bar z_4=0$ and defines the following function \ci{BPZ}
$$
{\cal G}^{12}_{43}(x,\bar x)=\lim_{z_1,\bar z_1\to \infty}
z_1^{2h_1} \bar z_1^{2\bar h_1}
\vev{O_{h_1,\bar h_1}(z_1,\bar z_1)O_{h_2,\bar h_2}(1,1)
 O_{h_3,\bar h_3}(x,\bar x)O_{h_4,\bar h_4}(0,0)}\,.
$$
One replaces the product of the last two operators according to \re{OPE},
calculates the resulting 3-point correlator using \re{3} and
applies the identity
$$
F(a;b;x\partial_\xi) (1-\xi)^{-c}\bigg|_{\xi=0}
=F(a,c;b;x)
$$
to express the function ${\cal G}^{12}_{43}$ in the following form
\be
{\cal G}^{12}_{43}=\sum_h
\Omega(h_1,h_2,h) \Omega(h,h_3,h_4) D^{-1}(h)
{\cal F}^{12}_{43}(h|x)
{\cal F}^{12}_{43}(\bar h|\bar x)\,,
\lab{s-ch}
\ee
where the notation was introduced for the conformal block
\be
{\cal F}^{12}_{43}(h|x)
=x^{h-h_3-h_4} F(h_3-h_4+h,h_2-h_1+h;2h;x)\,.
\lab{block}
\ee
The associativity of the operator algebra \re{OPE} implies the following
crossing symmetry condition \ci{BPZ}
\be
{\cal G}^{12}_{43}(x,\bar x)={\cal G}_{23}^{14}(1-x,1-\bar x)
=x^{-2h_3}\bar x^{-2\bar h_3}
{\cal G}^{42}_{13}\lr{\frac1{x},\frac1{\bar x}}\,.
\lab{dual}
\ee
Expression \re{s-ch} gives the $s-$channel expansion of the 4-point
correlation function over the partial waves corresponding to the
principal series representation of the $SL(2,\IC)$ group labelled by
the conformal weight $(h,\bar h)$. We recall however that the
representations $(h,\bar h)$ and $(1-h,1-\bar h)$ are unitary
equivalent and one has to combine together two terms in the sum \re{s-ch}
having the conformal weights $h$ and $1-h$. The $t-$ and $u-$channel
expansions of the same function follow immediately from the duality
property \re{dual}.

Let consider the special case, $h_1=h_2=1$ and $h_3=h_4=0$,
corresponding to the correlator of 4 reggeized gluon fields%
\footnote{To avoid singular contribution of the diagonal term in \re{OPE}
one has to consider the limit $h_{1,2}\to 1$ with $h_1\neq h_2$ rather than
put $h_1=h_2=1$.}
\baa
{\cal G}(x,\bar x)
&=& \lim_{z,\bar z \to \infty}
z^2 \vev{O_{1,0}(z,\bar z) O_{1,0}(1,1) O_{0,1}(x,\bar x) O_{0,1}(0,0)}
\\[3mm]
&=&\sum_h \Omega(h,1,1)\Omega(h,0,0) D^{-1}(h)
{\cal F}(h|x){\cal F}(\bar h|\bar x)
\,,
\eaa
with the conformal blocks
${\cal F}(h|x)=x^{h-2} F(h,h;2h;x)$ and
${\cal F}(\bar h|\bar x)=x^{\bar h} F(\bar h,\bar h;2\bar h;\bar x)$.
Following our interpretation of the transition amplitudes of the BFKL states
as correlators of quasiprimary fields $O_{h,\bar h}$ one has to compare
this expression with the 4-point Green function of the reggeized gluons
defined in \re{G}. We realize that the r.h.s.\ of \re{G}
is given by the product of conformal blocks
$$
G_{h,\bar h}(\rho_1,\rho_2;\rho_3,\rho_4)
=\frac{x^2}{B(1-h)} {\cal F}(h|x)  {\cal F}(\bar h|\bar x)
+(h \rightleftharpoons 1-h)
\,.
$$
Finally, using \re{11h}, \re{cc}, \re{D} and \re{norm} to calculate the product $\Omega(h,1,1)
\Omega(h,0,0)D^{-1}(h)$, one obtains the following relation
between the 4-point correlation function of quasiprimary operators and the
reggeon Green function
\be
{\cal G}(x,\bar x)=\frac{8\pi^2}{x^2}
\sum_h \lr{\Re\left[\psi(1)-\psi(h)\right]}^2
\left|\frac{h-\frac12}{h(h-1)}\right|^2
G_{h,\bar h}(\rho_1,\rho_2;\rho_3,\rho_4)\,.
\lab{str}
\ee
Although each term in this sum does not obey the crossing symmetry, the
symmetry is restored in their total sum due to \re{dual}.

We recognize the striking similarity of the expression \re{str}
and the BFKL scattering amplitude \re{Phi}. Namely, the second derivative
of $f_Y(\rho_1,\rho_2;\rho_3,\rho_4)$ with respect to rapidity
coincides up to $Y-$dependent factor with the 4-point correlation
function $x^2 {\cal G}(x,\bar x)$. To introduce the dependence on
the rapidity into the sum over conformal weights in \re{str} one notices
that the Green function $G_{h,\bar h}(\rho_1,\rho_2;\rho_3,\rho_4)$
diagonalizes the quadratic Casimir operator of the $SL(2,\IC)-$group,
$$
{\bf L}^2_{12}=z_{12}^2\partial_1\partial_2=h(h-1) \,,
$$
as well as the BFKL Hamiltonian ${\cal H}_{\rm BFKL}=
{\cal H}_{\rm BFKL}({\bf L}^2_{12})=\omega(h)$ with $\omega(h)$
being the energy of the BFKL states, \re{omega}.
This allows us to write the relation between the BFKL scattering amplitude
and the 4-point correlation function of reggeized gluon operators as
\be
{\partial^2_Y}
f_Y(\rho_1,\rho_2;\rho_3,\rho_4)=
\frac{(\as N_c)^2}{2\pi^4}
\e^{Y{\cal H}_{\rm BFKL}({\bf L}_{12}^2)}
x^2{\cal G}(x,\bar x)\,,
\lab{Eq}
\ee
where the energy of the BFKL state was replaced in \re{Phi}
by the BFKL Hamiltonian and anharmonic ratios, $x$ and $\bar x$,
were defined in \re{x}.

The following remarks are in order. The factor $\e^{Y{\cal H}_{\rm BFKL}}$
plays the role of the evolution operator with the rapidity interval $Y$
defining the $t-$channel evolution time of the system of two reggeons.
Although the
function ${\cal G}(x,\bar x)$ is crossing symmetric and it can be expanded
over the conformal blocks in different channel
using \re{s-ch} and \re{dual}, the BFKL kernel
${\cal H}_{\rm BFKL}({\bf L}_{12}^2)$ picks up the diagonal contribution
only in the channel $1+2\to 3+4$. As a result, the crossing symmetry of the
BFKL scattering amplitude $f_Y(\rho_1,\rho_2;\rho_3,\rho_4)$
is broken in the high-energy limit, $Y\gg 1$.

\section{Summary}

Studying the high-energy asymptotics of the scattering amplitude
of the perturbative hadronic states like heavy onia one is trying
to construct the effective $(1+2)-$theory which will describe the
effective QCD dynamics in the Regge limit. The main objects of the
effective theory are the $n=2,3,...$ reggeon compound states
which propagate in time defined as rapidity of the scattered particles
and interact on the 2-dimensional plane of impact parameters. The
corresponding bare interaction vertices are local in time and they can
be calculated by projecting the reggeon number changing kernels on
the wave functions of the compound states entering the vertex.

The reggeon transition kernels exhibit remarkable $SL(2,\IC)$ symmetry
suggesting that the interaction vertices of the reggeon compound states
in the Regge effective theory can be evaluated as correlators of
interpolating quasiprimary fields in two-dimensional conformal field
theory. In the present paper we have discussed this possibility by
calculating the interaction vertex of three BFKL states, $V(\alpha\to
\beta,\gamma)$.

Projecting the reggeon transition kernel $V_{(2,4)}$ on the wave
functions of three BFKL states entering the interaction vertex
and performing 2-dimensional integrations over the reggeon coordinates
we have obtained the analytical expression for the vertex
$V(\alpha\to\beta,\gamma)$ valid for arbitrary values of the conformal
weights of the BFKL states. For special values of the conformal
weights, $h_\alpha=h_\beta=h_\gamma=\frac12$, we have found the expression
for the bare triple BFKL Pomeron coupling which is of some importance
for Regge phenomenology as it enters into perturbative QCD description
of the inclusive cross sections of the diffractive dissociation of the
deep inelastic photon \ci{BW}.

We have shown that in the multi-color limit the vertex
$V(\alpha\to\beta,\gamma)$ satisfies bilinear relations
which were interpreted as associativity conditions of the
operator algebra of the operators $O_{h,\bar h}(z,\bar z)$
interpolating the BFKL states and defined as quasiprimary
fields in two-dimensional conformal field theory. We have constructed
the operator algebra of the fields $O_{h,\bar h}(z,\bar z)$
and demonstrated that their three-point correlation function
coincides with the interaction vertex of three BFKL states in
the multi-color limit.

\section*{Acknowledgments}

I am grateful to V.Braun for the discussions on numerical
calculation of the triple BFKL Pomeron vertex. I acknowledge helpful 
conversations with J.Bartels, C.Ewerz, A.Kaidalov and A.Mueller. 
I would also like to thank the Aspen Center for Physics where 
the initial part of this work was done.



\section*{Note added on November 14, 1997}

Recently I became aware that the same numerical expression for the planar 
contribution to the triple BFKL Pomeron vertex, \re{O1/2}, was obtained 
independently by different technique in the paper \ci{BNP}. 

\section*{Appendix A. Calculation of the nonplanar diagram}
\def\theequation{A.\arabic{equation}}

In this appendix we describe the calculation of the nonplanar diagram
\re{B}. Using the property \re{O} one can restrict analysis only to the 
positive values of the conformal spins $n_\alpha$, $n_\beta$, $n_\gamma > 0$ 
and rewrite \re{2-loop} as
\baa
\Lambda&=&
\int\!d^2 z_0\!\int\! d^2 z_1\, z_{01}^{n_\alpha+n_\beta+n_\gamma}
(\bar z_0\bar z_1)^{n_\alpha} ((1-\bar z_0)(1-\bar z_1))^{n_\beta}
\\
&&\times
|z_{01}^2|^{-2+\bar h_\alpha+\bar h_\beta+\bar h_\gamma}
|z_0^2 z_1^2|^{-h_\alpha}
|(1-z_0)^2 (1-z_1)^2|^{-h_\beta}.
\eaa
The calculation of the integral is based on the following
integral representation
\be
|z^2|^{-h}=\frac1{\Gamma(h)}\int_0^\infty d\alpha \,\alpha^{h-1}
\e^{-s z\bar z}\,.
\lab{alpha}
\ee
Repeatedly applying this formula to all $|z^2|-$factors we get additional
5-dimensional integral over the $\alpha-$parameters weighted with
$(\alpha_1\alpha_2)^{h_\alpha-1}(\alpha_3\alpha_4)^{h_\beta-1}
\alpha_5^{1-\bar h_\alpha-\bar h_\beta-\bar h_\gamma}$
and move all $|z^2|-$terms
into the exponent
$$
\exp\lr{-\alpha_1 |z_0^2|-\alpha_2|z_1^2|-\alpha_3|(1-z_0)^2|
-\alpha_4 |(1-z_1)^2|-\alpha_5|(z_0-z_1)^2|}\,.
$$
We note that integration goes over complex $z=x+iy$ and $\bar z=x-iy$
with the measure $d^2z=dxdy=\frac12dzd\bar z$.
One rotates the integration
contour over $y$ as $y\to iy$ without encountering any singularities
and integrates over $z$ and $\bar z$ along the real axis. Then, the
integration by parts over the holomorphic coordinates $z_0$ and $z_1$ gives
the product of two $\delta-$functions
$$
\delta(\alpha_1 \bar z_0-\alpha_3 (1-\bar z_0)+\alpha_5(\bar z_0-\bar z_1))\
\delta(\alpha_2 \bar z_1-\alpha_4 (1-\bar z_1)-\alpha_5(\bar z_0-\bar z_1))\,.
$$
Taking into account that the $\alpha-$parameters are positive, one can
satisfy the $\delta-$function constraints only for
$0\le \bar z_0 \le \bar z_1 \le 1$ or  $0\le \bar z_1 \le \bar z_0 \le 1$.
In the second case, one changes the integration variables $\bar z\to 1-\bar z$
and finds that the integral takes the same form as in the first case up
to a sign factor $(-)^{n_\alpha+n_\beta+n_\gamma}$. Finally, one
rescales $\alpha-$parameters to get rid of $\bar z-$dependent prefactors
in the arguments of $\delta-$functions and writes $\Lambda$ in the
form \re{form}
with the integrals over $\alpha$ and $\bar z$ factorized into
\baa
J &=& \int_0^\infty d\alpha_1 ... d\alpha_5\,
(\alpha_1\alpha_2)^{h_\alpha-1}(\alpha_3\alpha_4)^{h_\beta-1}
\alpha_5^{1-h_\alpha-h_\beta-h_\gamma}
\delta(\alpha_1-\alpha_3+\alpha_5)
\delta(\alpha_2-\alpha_4-\alpha_5)
\e^{-\alpha_3-\alpha_4}\,,
\\
\bar J &=& \int_0^1 d\bar z_0\int_0^{\bar z_0} d\bar z_1  \,
\bar z_{01}^{\bar h_\alpha+\bar h_\beta+\bar h_\gamma-2}
(\bar z_0\bar z_1)^{-\bar h_\alpha}
((1-\bar z_0)(1-\bar z_1))^{-\bar h_\beta}\,.
\eaa
Changing the integration variables in $J$ as $\alpha_1=\lambda u$,
$\alpha_4=\lambda(1-v)$ and $\alpha_5=\lambda(1-u+v)$ with
$0 \le \lambda < \infty$ and $0\le v \le u \le 1$ one gets
$$
J=\Gamma(h_\alpha+h_\beta-h_\gamma)
\int_0^1 du \int_0^u dv\, (u(1-u))^{h_\alpha-1}(v(1-v))^{h_\beta-1}
(u-v)^{1-h_\alpha-h_\beta-h_\gamma}\,.
$$
Applying the identity
\be
(1-v)^{h_\beta-1}=\sum_{k=0}^\infty v^k
\frac{\Gamma(1+k-h_\beta)}{\Gamma(k+1)\Gamma(1-h_\beta)}
\lab{tri}
\ee
one can express $J$ as an infinite sum of the ratio of $\Gamma-$functions
which can be summed into ${}_3F_2-$hypergeometric series
\baa
J=\frac{\Gamma(h_\alpha+h_\beta-h_\gamma)
       \Gamma(h_\beta)\Gamma(2-h_\alpha-h_\beta-h_\gamma)
        \Gamma(h_\alpha)\Gamma(1-h_\gamma)}
	{\Gamma(2-h_\alpha-h_\gamma)\Gamma(1+h_\alpha-h_\gamma)}
&&
\\
\times
{}_3F_2\left({{h_\beta, 1-h_\beta, 1-h_\gamma}\atop
   {2-h_\alpha-h_\gamma, 1+h_\alpha-h_\gamma}}
\bigg|1\right)\,.
\eaa
Calculating the integral $\bar J$ one applies the same identity to
$(1-\bar z_1)^{-\bar h_\beta}$ and obtains in similar way the
following result
\baa
\bar J=\frac{\Gamma(1-\bar h_\alpha)\Gamma(1-\bar h_\beta)
             \Gamma(-1+\bar h_\alpha+\bar h_\beta+\bar h_\gamma)
             \Gamma(-\bar h_\alpha+\bar h_\beta+\bar h_\gamma)}
	{\Gamma(\bar h_\beta+\bar h_\gamma)
	\Gamma(1-\bar h_\alpha+\bar h_\gamma)}
&&
\\
\times
{}_3F_2\left({{\bar h_\beta, 1-\bar h_\alpha,
   -\bar h_\alpha+\bar h_\beta+\bar h_\gamma}\atop
   {\bar h_\beta+\bar h_\gamma, 1-\bar h_\alpha+\bar h_\gamma}}
\bigg|1\right)\,.
\eaa
Thanks to the Whipple and Dixon identities \ci{B} the ${}_3F_2-$series
entering $J$ and $\bar J$ can be calculated as the ratio of $\Gamma-$functions
leading to the final result \re{Lam} for the function $\Lambda$.
We would like to stress that the same result can be obtained by using
the conventional technique of calculating 2-dimensional integrals in
conformal field theories \ci{DF}.

\section*{Appendix B. Calculation of the planar diagram}
\def\theequation{B.\arabic{equation}}

The calculation of the planar diagram is similar to that of the nonplanar
diagram. We choose the conformal spins of three BFKL states to be positive
and rewrite \re{Om} as
\baa
\Omega&=&\int d^2 z_0\int d^2 z_1\int d^2 z_2 \
(z_{01} \bar z_0 \bar z_1)^{n_\alpha}
(z_{12}(1-\bar z_1)(1-\bar z_2))^{n_\beta}
 z_{20}^{n_\gamma}
\\[3mm]
&& \times
 |z_{01}^2|^{\bar h_\alpha-1}
 |z_{12}^2|^{\bar h_\beta-1}
 |z_{20}^2|^{\bar h_\gamma-1}
 |z_0^2|^{-h_\alpha}|z_1^2|^{-h_\alpha}
 |(1-z_1)^2|^{-h_\beta}|(1-z_2)^2|^{-h_\beta}\,.
\eaa
Using the representation \re{alpha} one moves $|z^2|-$factors into the exponent
$$
\exp\lr{
-\alpha_1 |z_0^2|-\alpha_2 |z_1^2|-\alpha_3|(1-z_1)^2|-\alpha_4|(1-z_2)^2|
-\alpha_5 |z_{01}^2|-\alpha_6|z_{12}^2|-\alpha_7 |z_{20}^2|
}
$$
and integrates over the corresponding seven $\alpha-$parameters with the
weight $(\alpha_1\alpha_2)^{h_\alpha-1}(\alpha_3\alpha_4)^{h_\beta-1}$
$\alpha_5^{-\bar h_\alpha}\alpha_6^{-\bar h_\beta}\alpha_7^{-\bar h_\gamma}$.
Rotating the integration contours in complex $z-$plane and
integrating by parts over the holomorphic coordinates
$z_0$, $z_1$ and $z_2$ along the real axis one gets the product of three
$\delta-$functions
\be
\delta(\alpha_1 \bar z_0+\alpha_5 \bar z_{01}+\alpha_7 \bar z_{02})\
\delta(\alpha_2 \bar z_1+\alpha_3(\bar z_1-1)+\alpha_5\bar z_{10}+\alpha_6\bar
z_{12})\
\delta(\alpha_4(\bar z_2-1)+\alpha_6\bar z_{21}+\alpha_7\bar z_{20})\,,
\lab{df}
\ee
which restrict the possible values of antiholomophic coordinates $\bar z_0$,
$\bar z_1$ and $\bar z_2$. Namely, the $\delta-$function constraints define
three different regions:
\be
1{\rm st}:\quad 0 \le \bar z_0 \le \bar z_1 \le \bar z_2 \le 1\,,\quad
2{\rm nd}:\quad 0 \le \bar z_0 \le \bar z_2 \le \bar z_1 \le 1\,,\quad
3{\rm rd}:\quad 0 \le \bar z_1 \le \bar z_0 \le \bar z_2 \le 1\,.
\lab{reg}
\ee
In each of these regions one rescales the $\alpha-$parameters to get
rid of $\bar z-$dependent prefactors in \re{df} and factorizes $\alpha-$
and $\bar z-$ integrations into two integrals denoted as $J_a$ and
$\bar J_a$, respectively, with $a=1,\,2,\,3$ referring to the particular
region \re{reg}. This leads to the expression \re{Ja} for
the function $\Omega$ as a sum of three terms, in which
the integrals over $\alpha-$parameters corresponding to
three different regions are given by
$$
\lr{\begin{array}{c} J_1 \\[2mm]  J_2 \\[2mm] J_3 \end{array}}
=\int_0^\infty {\cal D}\alpha\,\e^{-\alpha_3-\alpha_4}\times
\lr{
\begin{array}{c}
 \delta(\alpha_1-\alpha_5-\alpha_7)
 \delta(\alpha_2-\alpha_3+\alpha_5-\alpha_6)
 \delta(-\alpha_4+\alpha_6+\alpha_7)
\\[2mm]
 \delta(\alpha_1-\alpha_5-\alpha_7)
 \delta(\alpha_2-\alpha_3+\alpha_5+\alpha_6)
 \delta(-\alpha_4-\alpha_6+\alpha_7)
\\[2mm]
 \delta(\alpha_1+\alpha_5-\alpha_7)
 \delta(\alpha_2-\alpha_3-\alpha_5-\alpha_6)
 \delta(-\alpha_4+\alpha_6+\alpha_7)
\end{array}
}
$$
with the integration measure
$$
\int_0^\infty {\cal D}\alpha\equiv
\int_0^\infty \prod_{j=1}^7 d\alpha_j\
(\alpha_1\alpha_2)^{h_\alpha-1}
(\alpha_3\alpha_4)^{h_\beta-1}
 \alpha_5^{-h_\alpha}
 \alpha_6^{-h_\beta}
 \alpha_7^{-h_\gamma}\,.
$$
The corresponding $\bar z-$integrals are given by
$$
\lr{\begin{array}{c} \bar J_1 \\[2mm]  \bar J_2 \\[2mm]  \bar J_3 \end{array}}
=\int_0^1 {\cal D}\bar z \times
\lr{
\begin{array}{l}
\theta(0 \le \bar z_0 \le \bar z_1 \le \bar z_2 \le 1)\ (-)^{n_\alpha+n_\beta}
\\[2mm]
\theta(0 \le \bar z_0 \le \bar z_2 \le \bar z_1 \le 1)\ (-)^{n_\alpha}
\\[2mm]
\theta(0 \le \bar z_1 \le \bar z_0 \le \bar z_2 \le 1)\ (-)^{n_\beta}
\end{array}
}
$$
where the $\theta-$functions define the integration region according
to \re{reg} and the integration measure is
$$
\int_0^1{\cal D}\bar z = \int_0^1 d\bar z_0\int_0^1 d\bar z_1\int_0^1 d\bar z_2
\ |\bar z_{01}|^{\bar h_\alpha-1}|\bar z_{12}|^{\bar h_\beta-1}
|\bar z_{20}|^{\bar h_\gamma-1}
\lr{\bar z_0\bar z_1}^{-\bar h_\alpha}\lr{(1-\bar z_1)(1-\bar z_2)}^{-\bar
h_\beta}\,.
$$
The integrals corresponding to the regions 2 and 3 are related to each other
as
$$
J_2(h_\alpha,h_\beta,h_\gamma)=J_3(h_\beta,h_\alpha,h_\gamma)\,,\qquad
\bar J_2(\bar h_\alpha,\bar h_\beta,\bar h_\gamma)
=\bar J_3(\bar h_\beta,\bar h_\alpha,\bar h_\gamma)\,.
$$
To obtain these relations one replaces the integration variables
$(\alpha_1,\alpha_2,\alpha_5)\rightleftharpoons (\alpha_4,\alpha_3,\alpha_6)$
and $(\bar z_0,\bar z_1,\bar z_2) \rightleftharpoons
(1-\bar z_2,1-\bar z_1,1-\bar z_0)$
in the expressions for $J_2$ and $\bar J_2$, respectively. In similar way,
one can show that $J_1$ and $\bar J_1$ are symmetric functions of
$h_\alpha$ and $h_\beta$
\be
J_1(h_\alpha,h_\beta,h_\gamma)=J_1(h_\beta,h_\alpha,h_\gamma)\,,\qquad
\bar J_1(\bar h_\alpha,\bar h_\beta,\bar h_\gamma)
=\bar J_1(\bar h_\beta,\bar h_\alpha,\bar h_\gamma)\,.
\lab{J2-J3}
\ee
The calculation of the integrals goes as follows. One first performs
the calculation of the functions $J_a$ and $\bar J_a$
for zero values of the conformal spins $n_\alpha=n_\beta=n_\gamma=0$,
at which the integrals entering these functions are convergent
and then analytically continues the final expressions to the
general values of the conformal weights.

\noindent
\underline{Integral $J_1$.}
One first replaces $\alpha_1=\alpha_5+\alpha_7$,
$\alpha_2=\alpha_3+\alpha_6-\alpha_5$ and $\alpha_4=\alpha_6+\alpha_7$
using the $\delta-$functions and then changes the remaining integration
variables as
$$
\alpha_3+\alpha_6+\alpha_7=\lambda\,,\qquad
\alpha_3+\alpha_6=\lambda x\,,\qquad
\alpha_3=\lambda x y\,,\qquad
\alpha_5=(\alpha_3+\alpha_6)(1-z)
$$
with $0\le \lambda < \infty$ and $0\le x\,, y\,, z  \le 1$. Integration
over $y$ and $z$ gives two ${}_2F_1-$hypergeometric functions leading to
\baa
J_1&=&\Gamma(h_\alpha+h_\beta-h_\gamma)\Gamma(1-h_\alpha)\Gamma(h_\alpha)
\Gamma(1-h_\beta)\Gamma(h_\beta)
\\
&&
\times\int_0^1 dx\, (1-x)^{-h_\gamma}
{}_2F_1(h_\alpha,1-h_\alpha;1;x)\, {}_2F_1(h_\beta,1-h_\beta;1;x)\,,
\eaa
which explicitly obeys \re{J2-J3}. For general values of the conformal
weights this integral can be expressed in terms of the Meijer's
$G-$function as follows. One uses the Mellin-Barnes representation
for one of the hypergeometric functions as an integral of $(1-x)^s$
with certain $\Gamma-$function prefactor along the $s-$contour
parallel to the imaginary axis and interchanges the order of $s-$ and
$x-$integration to obtain
$$
J_1=\Gamma(h_\alpha+h_\beta-h_\gamma)
\frac{\Gamma(h_\alpha)\Gamma(1-h_\alpha)}
{\Gamma(h_\beta)\Gamma(1-h_\beta)}\
G^{24}_{44}\lr{1 \bigg |
\begin{array}{l}
h_\beta, 1-h_\beta, h_\gamma, h_\gamma \\
0,0, -h_\alpha+h_\gamma, -1+h_\alpha+h_\gamma
\end{array}}\,.
$$

\noindent
\underline{Integral $J_2$.}
One integrates over $\alpha_1$, $\alpha_3$ and $\alpha_7$ using the
$\delta-$functions and replaces the integration variables as
$$
\alpha_2+\alpha_4+\alpha_5+\alpha_6=\lambda\,,\quad
\alpha_4+\alpha_5+\alpha_6=\lambda x_1\,,\quad
\alpha_4+\alpha_6=\lambda x_1 x_2\,,\quad
\alpha_4=\lambda x_1 x_2 x_3\,.
$$
with $0\le \lambda < \infty$ and $0\le x_{1,2,3} \le 1$. The integrals
over $x_i$ are factorized up to a factor $(1-x_1x_2x_3)^{h_\beta-1}$
which one expands using \re{tri} to obtain the expression for $J_1$ as
an infinite sum of the ratio $\Gamma-$functions. It can be summed into
${}_4F_3-$hypergeometric series as
\baa
J_2&=&\frac{
\Gamma(h_\alpha+h_\beta-h_\gamma)
\Gamma(1-h_\alpha)\Gamma(h_\alpha)
\Gamma(1-h_\beta)\Gamma(h_\beta)
\Gamma^2(1-h_\gamma)}
{\Gamma(1+h_\alpha-h_\gamma)\Gamma(2-h_\alpha-h_\gamma)}
\\
&&
\times{}_4F_3\left({{h_\beta, 1-h_\beta, 1-h_\gamma, 1-h_\gamma}\atop
   {1, 2-h_\alpha-h_\gamma, 1+h_\alpha-h_\gamma}}
\bigg|1\right)
\eaa

\noindent
\underline{Integral $\bar J_1$.}
The calculation of the integral is based on the identity
$$
(\bar z_2-\bar z_0)^{\bar h_\gamma-1}
=\sum_{k,n=0}^\infty \frac{(1-\bar h_\gamma)_{k+n}}{k!\, n!}
\bar z_0^k (1-\bar z_2)^n\,,
$$
where $(a)_k\equiv\Gamma(a+k)/\Gamma(a)$.
It allows to separate $\bar z_0-$ and $\bar z_2-$integrations and
obtain after simple calculation the expression for $\bar J_1$ in the form
of double series
$$
\bar J_1 = (-)^{n_\alpha+n_\beta}
\frac{\Gamma(\bar h_\alpha)\Gamma(\bar h_\beta)}
{\Gamma(1-\bar h_\gamma)}
\sum_{k,n=0}^\infty
\frac{\Gamma^2(1+k-\bar h_\alpha)\Gamma^2(1+n-\bar h_\beta)
\Gamma(1+k+n-\bar h_\gamma)}{\Gamma^2(1+k)\Gamma^2(1+n)
\Gamma(2+k+n-\bar h_\alpha-\bar h_\beta)}\,.
$$
One can check that the same series can be represented as the following
integral
\baa
\bar J_1 &=& (-)^{n_\alpha+n_\beta}
\frac{\Gamma(\bar h_\alpha)\Gamma(\bar h_\beta)
\Gamma^2(1-\bar h_\alpha)\Gamma^2(1-\bar h_\beta)}
{\Gamma(1-\bar h_\gamma)\Gamma(1-\bar h_\alpha-\bar h_\beta+\bar h_\gamma)}
\\
&&\times\int_0^1 dx\, x^{-\bar h_\gamma}
(1-x)^{-\bar h_\alpha-\bar h_\beta+\bar h_\gamma}
{}_2F_1(1-\bar h_\alpha,1-\bar h_\alpha;1;x)\,
{}_2F_1(1-\bar h_\beta,1-\bar h_\beta;1;x)\,,
\eaa
which obviously satisfies \re{J2-J3}. For general values of the conformal
weights this integral can be expressed in terms of the Meijer's $G-$function.
Repeating the same steps as for calculation of $J_1$ one obtains
$$
\bar J_1=
(-)^{n_\alpha+n_\beta}
\frac{\Gamma(1-\bar h_\alpha)
\Gamma(1-\bar h_\beta)}
{\Gamma(1-\bar h_\gamma)
\Gamma(1-\bar h_\alpha-\bar h_\beta+\bar h_\gamma)}
G^{33}_{44}\lr{1 \bigg |
\begin{array}{l}
\bar h_\beta, 1-\bar h_\beta, \bar h_\gamma, \bar h_\gamma \\
0,0, -\bar h_\alpha+\bar h_\gamma, -1+\bar h_\alpha+\bar h_\gamma
\end{array}}
$$

\noindent
\underline{Integral $\bar J_2$.}
Similar to the previous case, one applies the identity
$$
(\bar z_1-\bar z_0)^{\bar h_\alpha-1}
=\sum_{k,n=0}^\infty \frac{(1-\bar h_\alpha)_{k+n}}{k!\, n!}
\bar z_0^k (1-\bar z_1)^n
$$
to express the integral in the form of double series
$$
\bar J_2= (-)^{n_\alpha}
\frac{\Gamma(\bar h_\beta)\Gamma(\bar h_\gamma)}
{\Gamma(1-\bar h_\alpha)}
\sum_{k,n=0}^\infty
\frac{\Gamma^2(1+k-\bar h_\alpha)\Gamma^2(1+n-\bar h_\beta)
\Gamma(1+k+n)}{\Gamma^2(1+k)\Gamma^2(1+n)
\Gamma(2+k+n-\bar h_\alpha-\bar h_\beta+\bar h_\gamma)}\,.
$$
This series is given by
\baa
\bar J_2 &=& (-)^{n_\alpha}
\frac{\Gamma(\bar h_\beta)\Gamma(\bar h_\gamma)
\Gamma(1-\bar h_\alpha)\Gamma^2(1-\bar h_\beta)}
{\Gamma(1-\bar h_\alpha-\bar h_\beta+\bar h_\gamma)}
\\
&&\times\int_0^1 dx\,
(1-x)^{-\bar h_\alpha-\bar h_\beta+\bar h_\gamma}
{}_2F_1(1-\bar h_\alpha,1-\bar h_\alpha;1;x)\,
{}_2F_1(1-\bar h_\beta,1-\bar h_\beta;1;x)\,,
\eaa
Although this expression is not symmetric in $\bar h_\alpha$ and
$\bar h_\beta$ its asymmetry is contained in the ratio of $\Gamma-$functions.
Changing the integration variable as $y=\frac{x}{1-x}$ and using the
Mellin-Barnes representation one obtains the following expression
$$
\bar J_2=
(-)^{n_\alpha}
\frac{\Gamma(1-\bar h_\alpha)\Gamma(\bar h_\gamma)}
{\Gamma(\bar h_\beta)\Gamma(1-\bar h_\alpha-\bar h_\beta+\bar h_\gamma)}\
G^{42}_{44}\lr{1 \bigg |
\begin{array}{l}
\bar h_\beta, 1-\bar h_\beta, \bar h_\gamma, \bar h_\gamma \\
0,0, -\bar h_\alpha+\bar h_\gamma, -1+\bar h_\alpha+\bar h_\gamma
\end{array}}
$$

The Meijer's $G-$function entering into $J_a$ and $\bar J_a$
can be expressed in terms of the ${}_4F_3-$hyper\-geometric
series and their derivative with respect to parameters \ci{spec}.
We do not give here the explicit formulas but rather consider two
special physically most interesting cases.

\noindent
\underline{$h_\alpha=h_\beta=h$ and $h_\gamma=\frac12$}

This case corresponds to the coupling of two BFKL states with the
conformal weights $(h,\bar h)$ and the BFKL Pomeron. Simplification
occurs thanks to the Burchall-Chaundy identity \ci{B} which expresses
$\lr{{}_2F_1}^2$ as a sum of ${}_2F_1$. The values of integrals
are
\ba
J_1&=&\frac{\pi^2}2\frac{\cos^2(\pi h)}{\sin^2(\pi h)}
\frac{\Gamma(2h-\frac12)}{(h-\frac12)^2}
{}_6F_5(h)
\nonumber
\\
J_2&=&-\pi^2\frac{\cos(\pi h)}{\sin^2(\pi h)}
\frac{\Gamma(2h-\frac12)}{h-\frac12}
{}_4F_3(h)
\nonumber
\\
\bar J_1&=&-2\sqrt\pi \frac{\cos(\pi h)}{\sin(\pi h)}
\Gamma^2(1-\bar h)
\frac{\Gamma(2\bar h-\frac12)}{\bar h-\frac12}
{}_4F_3(\bar h)
\lab{JaP}
\\
\bar J_2&=&(-)^n \frac{\sqrt\pi}2
\frac{\cos^2(\pi h)}{\sin(\pi h)}
\Gamma^2(1-\bar h)
\frac{\Gamma(2\bar h-\frac12)}{(\bar h-\frac12)^2}
{}_6F_5(\bar h)
\nonumber
\ea
where we used the notations introduced in \re{FF}.
Substitution of these integrals into \re{Ja} yields the expression \re{hh1/2}
for the interaction vertex.

\noindent
\underline{$h_\alpha=h_\beta=1$ and $h_\gamma=h$}

This case corresponds to the coupling of two reggeized gluons to the
BFKL state with the conformal weight $(h,\bar h)$. Examining the integrals
one finds that each of them is separately divergent at
$h_\alpha=h_\beta=1$ (or equivalently, $\bar h_\alpha=\bar h_\beta=0$).
We consider instead the following values
$$
h_\alpha=1+i\nu_\alpha\,,\quad
h_\beta=1+i\nu_\beta
$$
and take the limit $\nu_\alpha$, $\nu_\beta\to 0$. Due to simplification
of the hypergeometric functions the integrals can be calculated as
\ba
J_1&=&-\frac{\Gamma(1-h)}{\nu_\alpha\nu_\beta}\left[
1+i(\nu_\alpha+\nu_\beta)\psi(1-h)+\CO(\nu^2)\right]
\nonumber
\\
J_2&=&-\frac{\Gamma(1-h)}{\nu_\alpha\nu_\beta}\left[
1+i\nu_\alpha\psi(1-h)+i\nu_\beta\psi(1)+\CO(\nu^2)\right]
\nonumber
\\
\bar J_1&=&\frac{(\bar h(\bar h-1))^{-1}}{\nu_\alpha\nu_\beta}
\left[-i(\nu_\alpha+\nu_\beta)
+2\nu_\alpha\nu_\beta\lr{\psi(\bar h)-\psi(1)}
\right.
\lab{JaB}
\\
&&\hspace*{25mm}
\left.+(\nu_\alpha+\nu_\beta)^2
\lr{\psi(\bar h+1)-\psi(1)-\frac1{\bar h-1}}
+\CO(\nu^3)
\right]
\nonumber
\\
\bar J_2&=&\frac{(\bar h(\bar h-1))^{-1}}{\nu_\beta}
\left[i+(\nu_\alpha+\nu_\beta)
\lr{\psi(1)-\psi(1+\bar h)+\frac1{\bar h-1}}
+\CO(\nu^2)
\right]
\nonumber
\ea
Combining these integrals together in the expression for the interaction
vertex we find that all terms singular at $\nu_\alpha$, $\nu_\beta\to 0$
cancel against each other giving a finite result \re{11h} for the vertex
$\Omega(1,1,h)$.

\bb{99}
\bi{L}    L.N. Lipatov, {\it Pomeron in quantum chromodynamics\/},
          in ``Perturbative QCD'', pp.411--489, ed. A.H. Mueller,
          World Scientific, Singapore, 1989; Phys. Rept. 286 (1997) 131.
\bi{Qua}  G.P. Korchemsky, Nucl. Phys. B462 (1996) 333.
\bi{BPZ}  A.A. Belavin, A.M. Polyakov and A.B. Zamolodchikov, Nucl. Phys. B241
          (1984) 333
\bi{BKP}  J. Bartels, Nucl. Phys. B175 (1980) 365;
\\        J. Kwiecinski and M. Praszalowicz, Phys. Lett. B94 (1980) 413.
\bi{FK}   L.D. Faddeev and G.P. Korchemsky, Phys. Lett. B342 (1995) 311.
\bi{LL}   L.N. Lipatov, JETP Lett. 59 (1994) 596.
\bi{B}    J. Bartels,  Z.Phys. C60 (1993) 471; Phys. Lett. B298 (1993) 204.
\bi{BW}   J. Bartels and M. Wusthoff, Z.Phys. C66 (1995) 157.
\bi{BLW}  J. Bartels, L.N. Lipatov and M. Wusthoff, Nucl. Phys. B464 (1996) 298.
\bi{BE}   C. Ewerz, preprint DESY-97-130, Jul. 1996 [hep-ph/9707257].
\bi{Lot}  H. Lotter, Doctoral Thesis, preprint DESY-96-262, Dec. 1996
          [hep-ph/9705288].
\bi{P}    A.M. Polyakov, ZhETF 66 (1974) 23.	  
\bi{RP}   R. Peschanski, Phys. Lett B409 (1997) 491; preprint 
          Saclay-SPhT-97-034, Apr. 1997 [hep-ph/9704342].
\bi{M}    A.H. Mueller, Nucl. Phys. B437 (1995) 107.
\bi{spec} {\it Higher transcendental functions\/},
          ed. A. Erd\'elyi, McGraw-Hill, 1953.
\bi{VB}   I.I. Balitsky and V.M. Braun, unpublished.
\bi{DF}   V.S. Dotsenko and V.A. Fateev, Nucl. Phys. B240 (1984) 312;
          B251 (1985) 691.
\bi{BNP}  A. Bialas, H. Navelet and R. Peschanski, Saclay-T97-131, Nov. 1997,
          [hep-ph/9711236] (revised version) and in preparation.

\eb
\end{document}